\documentclass[prd,print,
superscriptaddress,showpacs,nofootinbib,%
tightenlines,onecolumn
]{revtex4}
\usepackage{natbib}
\usepackage{epsfig}
\usepackage{color}
\usepackage{slashed}
\usepackage{amssymb}
\usepackage{amsmath}

\newcommand{\ben}{\begin{displaymath}}
\newcommand{\een}{\end{displaymath}}
\newcommand{\be}{\begin{equation}}
\newcommand{\ee}{\end{equation}}
\newcommand{\bea}{\begin{eqnarray}}
\newcommand{\eea}{\end{eqnarray}}

\begin{document}

\title{Towards baryon-baryon scattering in manifestly Lorentz-invariant \\ formulation of SU(3) baryon chiral perturbation theory}
\author{V.~Baru}
\affiliation{Helmholtz-Institut f\"ur Strahlen- und Kernphysik and Bethe Center
for Theoretical Physics, Universit\"at Bonn, D-53115 Bonn, Germany}
\affiliation{Institute for Theoretical and Experimental Physics, B. Cheremushkinskaya 25, 117218 Moscow, Russia}
\affiliation{P.N. Lebedev Physical Institute of the Russian Academy of Sciences, 119991, Leninskiy Prospect 53, Moscow, Russia}
\author{E.~Epelbaum}
 \affiliation{Ruhr University Bochum, Faculty of Physics and Astronomy,
Institute for Theoretical Physics II, D-44870 Bochum, Germany}
\author{J.~Gegelia}
\affiliation{Ruhr University Bochum, Faculty of Physics and Astronomy,
Institute for Theoretical Physics II, D-44870 Bochum, Germany}
 \affiliation{Tbilisi State  University,  0186 Tbilisi,
 Georgia}
\author{X.-L.~Ren}
 \affiliation{Ruhr University Bochum, Faculty of Physics and Astronomy,
Institute for Theoretical Physics II, D-44870 Bochum, Germany}

\date{6 May, 2019}

\begin{abstract}
We study baryon-baryon scattering by applying time-ordered
perturbation theory to the manifestly Lorentz-invariant formulation
of SU(3) baryon chiral perturbation theory. We derive the
corresponding diagrammatic rules 
paying
special attention to complications caused by momentum-dependent
interactions and propagators of particles with non-zero spin.
We define the effective potential as a sum of
two-baryon irreducible contributions of time-ordered diagrams and
derive a system of integral equations for the
scattering amplitude, which provides a coupled-channel 
generalization of the Kadyshevsky equation.
The obtained leading-order baryon-baryon potentials are perturbatively
renormalizable, and the corresponding integral equations have unique
solutions in all partial waves.  
We discuss the issue of additional finite subtractions required to
improve the ultraviolet convergence of (finite) loop integrals on the
example of nucleon-nucleon scattering in the $^3P_0$ partial wave. 
Assuming that corrections beyond leading order can be treated perturbatively, we
obtain a fully renormalizable  formalism which can be employed to study
baryon-baryon 
scattering.

\end{abstract}

\maketitle

\section{\label{introduction}Introduction}

Nuclear systems with non-vanishing strangeness provide an important
connection between nuclear, particle and astrophysics.
Hypernuclei emerging from replacing one or several nucleons in a nucleus
by hyperons serve as a testing ground for effects of strange quarks in
nuclear matter. In addition to the nuclear forces, hyperon-nucleon
(YN) interactions play a crucial role for understanding hypernuclear binding.
Experiments with hypernuclei aiming at the determination of the YN,
hyperon-hyperon (YY) and cascade-nucleon ($\Xi$N) interactions 
are carried out at various laboratories world-wide such as CERN,
DA$\Phi$NE, GSI, JLab, J-PARC, KEK, MAMI, RHIC and will also be
performed at the future FAIR facility, see
Refs.~\cite{Pochodzalla:2011rz,Esser:2013aya,Feliciello:2015dua,Gal:2016boi}
for review articles of hypernuclear physics. 

Given that experiments involving hypernuclei are rather challenging,
a particularly valuable source of information on YN  and YY
interactions is provided by lattice QCD.
Several lattice-QCD groups including HAL QCD
\cite{Doi:2017zov,Nemura:2017vjc,Sasaki:2018mzh} and NPLQCD
\cite{Beane:2013br,Beane:2012ey,Beane:2012vq,Beane:2010em,Beane:2008dv}
Collaborations, Yamazaki et
al.~\cite{Yamazaki:2012hi,Yamazaki:2015asa}
and the Mainz group \cite{Hanlon:2018yfv} 
have already 
produced interesting results on baryon-baryon (BB) systems,
and more results will become available in
the near future.
Although some lattice simulations are already approaching the physical values of the light quarks,  most of the available lattice QCD calculations 
still correspond to unphysically large values of quark masses and require a reliable
theoretical framework to perform extrapolations to their physical values.

Chiral effective field theory (ChEFT) for few-baryon systems offers a
natural approach to analyze low-energy properties of (hyper)nuclei and
to perform chiral extrapolations. It goes back to  the
seminal papers by Weinberg, who proposed a way of extending chiral
perturbation theory to systems involving two and more nucleons
\cite{Weinberg:rz,Weinberg:um}.
In the resulting ChEFT approach  
the power counting rules
are applied to the effective BB potential defined
as a sum of contributions of two-baryon-irreducible diagrams.
The scattering amplitude  is then obtained by solving the
Lippmann-Schwinger (LS) or Schr\"odinger equations.
For reviews of ChEFT in the few-body sector see
Refs.~\cite{Bedaque:2002mn,Epelbaum:2005pn,Epelbaum:2008ga,Machleidt:2011zz,Epelbaum:2012vx}. 

The Bonn-J\"ulich and later Munich groups have pioneered chiral EFT
for BB interactions in the strange sector using the non-relativistic
formulation
\cite{Meissner:2016ood,Haidenbauer:2015zqb,Petschauer:2015nea,Haidenbauer:2014rna,Haidenbauer:2013oca,Haidenbauer:2011za,Haidenbauer:2011ah,
Haidenbauer:2009qn,Polinder:2007mp,Haidenbauer:2007ra,Polinder:2006zh,Haidenbauer:2018gvg,Haidenbauer:2014uua,Haidenbauer:2017sws}.
In these studies  the standard Weinberg
power counting for nucleon-nucleon (NN) interactions extended to the
strangeness-$S = -1$  (i.e.~$\Lambda$N, $\Sigma$N) and strangeness-$S
= -2$  (i.e.~$\Lambda\Lambda$, $\Sigma\Sigma$ and $\Xi$N) systems is utilized. 
At leading order (LO) in the Weinberg power counting,  the two-baryon
potentials consist of four-baryon contact interactions without
derivatives and the one-pseudoscalar-meson
exchange.
At next-to-leading order (NLO) one has to take into account the
contributions from two-pseudoscalar-meson exchange diagrams and from
four-baryon 
contact interactions with two derivatives or a single insertion of the
quark masses.
To regularize the ultraviolet divergences appearing from iterations
of the LS equation already at leading order, a
finite-cutoff regularitazion has been employed using 
exponential cutoffs in the range of $500\ldots
700$~MeV.

A modified  approach to 
NN scattering has been proposed in
Ref.~\cite{Epelbaum:2012ua}. This novel framework employs time-ordered
perturbation theory (TOPT) and
relies on the manifestly Lorentz-invariant effective Lagrangian. It 
offers a perturbatively 
renormalizable modification of Weinberg's approach and has already
been explored in the non-strange sector \cite{Epelbaum:2013ij,Epelbaum:2013naa,Epelbaum:2015sha}. 
Dirac spinors have been kept in their full form and an alternative power counting has been suggested in Ref.~\cite{Ren:2016jna}.
First applications of the formalism based on the Lorentz-invariant Lagrangian to BB systems with non-zero strangeness
can be found in
Refs.~\cite{Li:2016mln,Ren:2018xxd,Li:2016paq,Li:2018tbt,Song:2018qqm}.\footnote{The
  first application of the  modified Weinberg  approach to hadronic
  molecules can be found in Ref.~\cite{Baru:2015tfa}.}
In these exploratory studies, the BB scattering amplitudes
were obtained by solving  the (generalized) Kadyshevsky equation \cite{kadyshevsky}, 
however, a {\it systematic} approach to the SU(3) sector using TOPT,
which would allow for a straightforward generalization beyond the
leading order,
has not been formulated yet. In this paper we fill this gap by generalizing the
modified Weinberg approach of Ref.~\cite{Epelbaum:2012ua} to the
SU(3) sector and work out in detail the TOPT diagrammatic rules
for particles with non-zero spin and interactions involving time derivatives.
To achieve this goal, we start with the Lorentz-invariant effective
Lagrangian and derive the diagrammatic rules of TOPT by integrating 
over zeroth components of loop momenta in Feynman diagrams for BB scattering. 
Special care is taken to deal with complications caused by momentum-dependent
interactions and propagators of particles with non-zero spin. We
provide details which are not given in
Ref.~\cite{Epelbaum:2012ua} and following papers. 
The obtained rules of TOPT can be applied systematically to all orders
in the loop expansion.
Using  the standard Weinberg
power counting for diagrams contributing to BB
scattering, one has to take into account an infinite number of graphs
already at LO. We define the effective potential as a sum of all
possible two-baryon-irreducible TOPT
diagrams and obtain the scattering amplitudes by solving the
corresponding integral equations. The resulting formulation permits a
systematic investigation of few-baryon systems using both a
renormalizable approach, which relies on a perturbative treatment of
corrections beyond LO and allows one to completely eliminate the
ultraviolet cutoff, and a conventional scheme based on iterating a truncated
potential to all orders, which requires the ultraviolet cutoff to be
chosen  of the order of the hard scale of the problem
\cite{Lepage:1997cs,Gegelia:1998iu,Park:1998cu,Lepage:1999kt,Epelbaum:2004fk,Gegelia:2004pz,Epelbaum:2006pt,Epelbaum:2018zli}.
In the latter case, the manifestly Lorentz-invariant formulation of
chiral EFT is expected to permit a larger cutoff variation as compared
to the conventional non-relativistic framework, which is especially
important for the SU(3) sector. 

Our paper is organized as follows: in section~\ref{effective_Lagrangian} we
work out  the rules of TOPT for a system of baryons interacting with pseudoscalar
mesons including momentum-dependent vertices. A system of integral
equations for BB scattering is derived 
in section~\ref{intequation}. 
Next, in section~\ref{Calculations} we discuss the LO BB potential and the renormalization. 
The results of our work are
summarized in  section~\ref{conclusions}.

\section{Diagrammatic rules in time-ordered perturbation theory}
\label{effective_Lagrangian}

To formulate the theoretical framework describing BB
scattering in SU(3) baryon chiral perturbation theory (BChPT) by applying the rules of TOPT
we start with the manifestly Lorentz-invariant effective Lagrangian. 
It consists of the purely mesonic, single-baryon, two-baryon, $\ldots$ parts,
\begin{equation}
{\cal L}_{\rm eff}={\cal L}_{\rm \phi}+{\cal L}_{{\rm \phi B}}+{\cal
L}_{\rm BB}+\cdots. \label{inlagr}
\end{equation}
The effective Lagrangian is
organised as an expansion in powers of the quark masses and derivatives.
{ The lowest-order mesonic Lagrangian can be found in Ref.~\cite{Gasser:1984yg}.
The lowest-order Lagrangian in the single-baryon sector is given by
\begin{equation}
{\cal L}_{{\rm \phi B}}^{(1)}=\mbox{Tr} \left\{ \bar {\rm B} \left( i\gamma_\mu D^\mu -m \right)  {\rm B} \right\}  +\frac{D/F}{2}
\mbox{Tr} \left\{\bar {\rm B} \gamma_\mu \gamma_5 [u^\mu,{\rm B}]_{\pm}\right\}  ,
\label{lolagr}
\end{equation}
where $D$ and $F$ are coupling constants,  $D_\mu{\rm B} = \partial_\mu {\rm B}+[[u^\dagger,\partial_\mu u], {\rm B}] $
denotes the covariant derivative with
$u_\mu =iu^{\dagger} \partial_\mu U u^{\dagger},u^2=\exp\left(\sqrt{2}\, i P/ F_0 \right).$ 
 Next, $P$ and $B$ are the  irreducible 
 octet representations of ${\rm SU}(3)_f$ for the Goldstone bosons and  baryons, respectively (see, e.g. Ref.~\cite{Petschauer:2015nea}),
and $F_0$ is the meson decay constant in the chiral
limit and we consider the
isospin-symmetric case. 

}

    The effective BB Lagrangian contains terms with an increasing number of derivatives acting on the baryon field.
Using field redefinitions and re-organizing certain terms one can achieve that the four-baryon effective Lagrangian contributing to the
LO BB potential involves only the following terms without derivatives 
\cite{Polinder:2007mp}:
\begin{equation}
{\cal L}_{\rm BB}^{(0)} = C^1_i \, \mbox{Tr} \left\{ \bar {\rm B}_\alpha \bar {\rm B}_\beta \left(  \Gamma_i {\rm B}\right)_\beta \left(  \Gamma_i {\rm B}\right)_\alpha  \right\}
+C^2_i \, \mbox{Tr} \left\{ \bar {\rm B}_\alpha  \left(  \Gamma_i {\rm B}\right)_\alpha \bar {\rm B}_\beta \left(  \Gamma_i {\rm B}\right)_\beta  \right\}  
+ C^3_i \, \mbox{Tr} \left\{ \bar {\rm B}_\alpha  \left(  \Gamma_i {\rm B}\right)_\alpha  \right\} \mbox{Tr} \left\{ \bar {\rm B}_\beta \left(  \Gamma_i {\rm B}\right)_\beta  \right\}  ,
\label{NNLagrdreg}
\end{equation}
where $C_i^1$, $C_i^2$ and  $C_i^3$ are coupling constants,  $\alpha$ and $\beta$ are the Dirac spinor indices and
$
\Gamma_1=1, \ \ \  \Gamma_2=\gamma^\mu, \ \ \  \Gamma_3=\sigma^{\mu\nu}, \ \ \  \Gamma_4=\gamma^\mu\gamma_5, \ \ \  \Gamma_5=\gamma_5 .
$
Notice that the $ \Gamma_5$-term yields a vanishing contribution at LO,
i.e.~it starts contributing at NLO.

Our aim is to obtain diagrammatic rules of 
TOPT.
Notice that the systematic technique of Ref.~\cite{Sterman:1994ce}
leads to the TOPT rules, which 
are not directly applicable to vertices involving time components of momenta.
Additional  complications emerge from treating particles with non-zero
spin. In particular,  zeroth components
of momenta in the numerators of propagators of particles with non-zero
spin cannot be substituted by their on-shell values. To our surprise, we were
unable to find in the literature a more complete treatment of TOPT
applicable to the case at hand. 

To obtain the rules of TOPT  we follow the usual procedure by first
drawing all possible Feynman diagrams (in principle, an 
infinite number of them) 
relevant for the process of interest, assigning the momenta to
propagators associated with internal lines and performing the trivial
momentum integrations using the delta functions appearing at the  vertices. The
remaining overall delta function ensures momentum
conservation for the external legs of a diagram. Next, we perform
integrations over the zeroth components of the loop momenta. This
leads to a decomposition of each Feynman diagram into a sum of
time-ordered diagrams.

\begin{figure}
\includegraphics[width=0.6\textwidth]{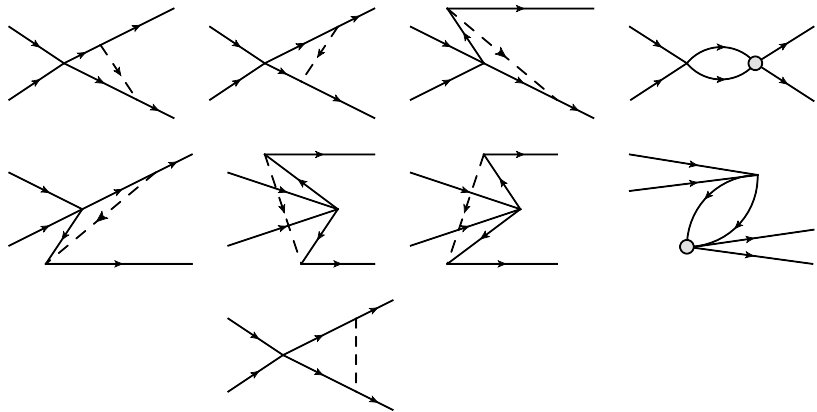}
\caption[]{\label{Triangle} One-loop diagrams contributing to the NN scattering. The first two rows represent eight time-ordered diagrams
corresponding to one Feynman diagram shown in the third row. The solid and dashed lines correspond to the nucleons and pions, respectively. 
The contact interaction with filled blob represents the additional
vertex due to the momentum-dependent pion-nucleon interaction as
explained in the text.}
\end{figure}

To demonstrate the above procedure we consider a Feynman diagram
contributing to the NN scattering amplitude and depicted in
the third line of Fig~\ref{Triangle}. 
Omitting SU(3) coefficients, the corresponding expression   reads  
\begin{equation}
\frac{g_A^2}{4 F_0^2} \int \frac{d^4k}{(2 \pi)^4} \frac{ \bar u_\alpha(p_3)\left[ \gamma_5 k\hspace{-.5em}/\hspace{.1em}\right]_{\alpha\lambda} (p_3\hspace{-.85em}/\hspace{.1em}+k\hspace{-.5em}/\hspace{.1em}+m_N)_{\lambda\beta} C_{\beta\delta,\mu\nu}
\bar u_\gamma(p_4) \left[ \gamma_5 k\hspace{-.5em}/\hspace{.1em}\right]_{\gamma\sigma} (p_4\hspace{-.85em}/\hspace{.1em} - k\hspace{-.5em}/\hspace{.1em}+m_N)_{\sigma\delta}\,
u_\mu(p_1) u_\nu(p_2)
}{\left( k^2-M_\pi^2 + i \epsilon \right)\left( (p_3+k)^2-m_N^2 + i \epsilon \right)\left( (p_4-k)^2-m_N^2 + i \epsilon \right)},
\label{TrFDiagram}
\end{equation}
where $g_A=D+F$ is the axial coupling, $p_1$ and $p_2$ ($p_3$ and $p_4$) are the four-momenta of
incoming (outgoing) nucleons { with $p_i^2=m_N^2$ and $C_{\beta\delta,\mu\nu}$ stands for
the four-nucleon interaction with the Dirac-spinor indices $\beta\delta,\mu\nu$.} 

In our calculations we use the Dirac spinors with the four-momentum $q$ 
\be\label{nuclspin}
 u(q) = \left(\frac{\omega(q,m_N)+m_N}{2m_N}\right)^{1/2}    \left(\begin{array}{c} \chi \\
                        \frac{\vec{\sigma}\cdot\vec{q} \ \chi }{\omega(q,m_N)+m_N}
                 \end{array}\right), 
\ee 
where $\omega(q,M):=\left(\vec q\,{  }^2+M^2\right)^{1/2}$, 
$\chi$ is the two-component spinor,
{ and for the nucleon propagator we apply
\begin{equation}
(p\hspace{-.45em}/\hspace{.1em}+m_N)_{\mu\nu} \equiv 2 \, m_N \sum \,u_\mu(p) \bar u_\nu(p)
  + \left[ p_0- \omega(p, m_N)\right] [\gamma_0]_{\mu\nu}\,,
\label{Sfpexpanded}
\end{equation}
where the summation is done over polarizations.   
 Next, we make use of the following identities  for the propagator with the momentum $q$ and the mass $m$
\begin{eqnarray} \label{denrewrite}
\frac{1}{q^2-m^2+i\epsilon}&=&\frac{1}{2 \,\omega(q,m)}\left( \frac{1}{q^0-\omega(q,m)+i\epsilon}+\frac{1}{-q^0-\omega(p,m)+i\epsilon}\right),\nonumber\\[2mm]
 \frac{q^0}{q^2-m^2+i\epsilon}&=&\frac{1}{2 \,\omega(q,m)}\left( \frac{\omega(q,m)}{q^0-\omega(q,m)+i\epsilon}+\frac{-\omega(q,m)}{-q^0-\omega(p,m)+i\epsilon}\right),\\[2mm]
 \frac{{(q^0)}^2}{q^2-m^2+i\epsilon}&=&1+ \frac{1}{2 \,\omega(q,m)}\left( \frac{\omega(q,m)^2}{q^0-\omega(q,m)+i\epsilon}+\frac{(-\omega(q,m))^2}{-q^0-\omega(p,m)+i\epsilon}\right)\nonumber.
\end{eqnarray}
These relations turn out to be particularly useful for deriving the diagrammatic rules of  TOPT. 
Then, 
we apply the relations given by Eqs.~\eqref{denrewrite} to the pion propagator and simplify the expression for the amplitude by collecting terms with equal denominators.
Those terms which have all poles on the same side of the real axis in
the complex $k_0$-plane  vanish after the integration over $k_0$ as a consequence of the Cauchy theorem.  For
the remaining terms, closing the contour of integration on the sides with single poles
and picking up the corresponding residues we obtain eight
contributions which can be represented as time-ordered diagrams shown in first two lines in Fig~\ref{Triangle}.
We stress that due to the cancellation of the $k_0$-dependent vertices
with the pion propagator, see the first term (i.e.~the unity) on the right-hand side of the last line of Eq.~(\ref{denrewrite}),
the Feynman diagram shown in Fig.~\ref{Triangle} contains also  purely short-range TOPT contributions. 
It is however  important to understand that a set of TOPT contributions, as shown by the diagrams in Fig.~\ref{Triangle}, is not unique and can be identically regrouped 
in such a way that the individual contributions are different but their sum is the same.  
Indeed, instead of applying Eqs.~\eqref{denrewrite} to the pion propagator  we could first use the identity  $k^0 = (p_3^0+k^0)- p_3^0$ and then use Eqs.~\eqref{denrewrite} for the nucleon 
propagators, which would result in different topologies. In particular, no purely short-range TOPT contributions  would  emerge in this case.  While  
in general the choice of TOPT topologies is just a matter of convenience, in some cases it is beneficial to  employ Eqs.~\eqref{denrewrite}  to the nucleon propagator to  separate the irreducible 
contribution (a contribution which does not possess the NN cut) from the Feynman diagram, see e.g. Ref.~\cite{Lensky:2005jc}  where this trick was applied to identify the irreducible one-loop diagrams contributing to the reaction $NN\to NN\pi$.

 For the sake of compactness, we do not provide the explicit TOPT expressions corresponding to the amplitude given by Eq.~\eqref{TrFDiagram} 
 but rather formulate the  general  rules of TOPT for  processes with
baryons and pseudoscalar mesons  
in the initial and final states.
\footnote{Modifications are
  required for processes involving anti-baryons in initial and final
  states.}
In the absence of fermion fields and interaction terms with time
derivatives, the rules given below reduce to the ones of
Ref.~\cite{Weinberg1}.

\noindent
The $S$ matrix for a process $\alpha\to\beta$ can be written as
\begin{eqnarray}
S_{\beta\alpha}&=& \delta_{\beta\alpha}-(2 \pi)^4
i\,M_{\beta\alpha}\delta^4(P_\beta-P_\alpha) 
\,  \Pi_{B}^{\alpha,\beta}\frac{(2
\pi)^{-3/2}}{(2 \omega(p_B,m_B))^{1/2}} \, \Pi_{F}^{\alpha,\beta} (2
\pi)^{-3/2}\left(\frac{m_F}{\omega(p_F,m_F)}\right)^{1/2},
\label{Smatrix}
\end{eqnarray}
where $P^\mu$ is the total four-momentum
and $\Pi_{B/F}^{\alpha,\beta}$ denotes a product of the expressions following these symbols over all bosons (labelled by B)/fermions (labelled by F) in
the initial and final states. The  invariant amplitude $M_{\beta\alpha}$ is obtained by
using the following diagrammatic rules:

\begin{itemize}
\item
Draw all possible time-ordered diagrams for the process 
$\alpha\to\beta$.\footnote{Do not include diagrams with closed baryon lines - for low-energy processes their contributions are taken into account by redefining coupling constants of the effective Lagrangian.} 
That is, draw each Feynman diagram with $N$ vertices
$N!$ times while ordering the vertices in every possible way in a sequence
running from right to left.  Label each line with a four-momentum
$p=(p_0,\vec p)$ as prescribed by the corresponding Feynman
diagram.
\item
Include a factor
\begin{equation}
(2 \,\omega(p_i,M_i))^{-1}
\label{efactor}
\end{equation}
for every internal line corresponding to a pseudoscalar meson with the mass $M_i$ and four-momentum $p_i$.
\item
For every internal baryon line with the momentum $p_j$ and mass $m_j$
include a factor
\begin{equation}
\frac{m_j}{\omega(p_j,m_j)}\sum \,u(p_j) \bar u(p_j)\,,
\label{bnum}
\end{equation}
where summation is done over polarizations.
\item
For every internal anti-baryon line (i.e. the line with baryon number flowing opposite to the time direction) with the momentum $p_i$ and mass
$m_i$ include a factor
\begin{equation}
\frac{m_i}{\omega(p_i,m_i)}\sum \,u(p_i) \bar u(p_i) - \gamma_0 \,,
\label{abnum}
\end{equation}
where summation is done over polarizations.
\item
  For every incoming (outgoing) external fermion with momentum $p$
  ($p'$) include $u(p)$ ($\bar u(p')$).
\item
For interaction vertices, use the ones of the standard Feynman rules.
Care has to be taken of zeroth components of momenta appearing in the vertices.
Indeed, as follows from Eq.~(\ref{denrewrite}),
for interactions containing one power of the zeroth component of a
momentum, this zeroth component has to be replaced by the  energy 
 of the particle carrying this momentum if it corresponds to a 
 particle and  
 by $-1$ times the energy  if the line corresponds to an 
 antiparticle. 
 If a Feynman diagram involves two vertices containing zeroth
 component of a momentum corresponding to the same line, the rule of the previous sentence applies to each of the vertices. 
 However, in addition, one has to include also the TOPT diagrams 
 where the two vertices and the
 connecting propagator cancel each other thus resulting in an effective contact contribution -- see the unity in the very last line of Eq.(\ref{denrewrite}). 
  We also note that interaction terms containing two or more time derivatives acting on the same field are always eliminated from the effective Lagrangian by using suitable field redefinitions.  
\item
For every intermediate state, i.e.~a set of lines between any
two vertices corresponding to particles enumerated with $\gamma$, include an energy denominator
\begin{equation}
\left[E -\sum_\gamma \omega(p_\gamma,m_\gamma)+i\,\epsilon\right]^{-1},
\label{denominator}
\end{equation}
where $E$ is the total energy of the system, i.e.~the sum of the energies of all particles in initial/final state.
\item
  Integrate over all internal 
momenta ($\int d^3 \vec k_i/(2\pi)^3$ for each $k_i$), and add
together the
contributions from all time-ordered diagrams.
\end{itemize}

\section{ Integral equations for  baryon-baryon scattering }
\label{intequation}

The BB scattering amplitude $M_{BB}$ is obtained from the four-point
vertex function $\tilde\Gamma_{4B}$ by applying the standard LSZ
formula
\begin{equation}
M_{BB}=Z_B^2\, \bar u(p_3) \bar u(p_4)\,\tilde\Gamma_{4B}\,u(p_1) u(p_2) \equiv Z_B^2\, \tilde {\cal M},
\label{LSZtosh}
\end{equation}
where $Z_B$ is the residue of the dressed baryon propagator and $u$, $\bar u$
are Dirac spinors corresponding to the incoming and outgoing baryons. 
The on-shell amplitude $\tilde {\cal M}$  is given as a sum of an
infinite number of TOPT diagrams. Notice that it does not include
diagrams with corrections on external legs.  

 For BB scattering, the purely two-baryon intermediate
 states are enhanced \cite{Weinberg:rz}. Therefore, it is convenient to 
define the effective potential as a sum of all possible two-particle
irreducible TOPT diagrams. 
The amplitude $\tilde {\cal M}$  is then given by an infinite series
\begin{eqnarray}
\tilde {\cal M} = \tilde V+\bar V G\, \bar V +\bar V G\, V G\, \bar V+ \bar V G\, V G\, V G\, \bar V +\cdots
= \tilde V+\bar V G\, \bar V +\bar V G\,{\cal M} G\, \bar V,
\label{Tseries}
\end{eqnarray}
where $G$ is the two-baryon Green's function and $\tilde {\cal M}$,
${\cal M}$, $\tilde V$, $\bar V$ and $V$  are the on-shell amplitude, 
the off-shell amplitude, the on-shell potential, the half-off-shell
potential and the off-shell potential,   respectively. The on-shell
potential $\tilde V$ does not include diagrams with corrections on
external legs. The half-off-shell potential $\bar V$ does not include
diagrams with corrections on external legs with on-shell momenta while
the off-shell potential $V$ also includes diagrams with corrections on
external legs.   

The off-shell amplitude ${\cal M}$ can be obtained by solving the following equation:
\begin{equation}
{\cal M}=V+ V G\,{\cal M}\,.
\label{Teq1}
\end{equation}
To cover all different processes with strangeness $S=0,-1,-2$ in the
isospin limit, Eq.~(\ref{Teq1}) has to be understood as a $7\times 7$ matrix equation,
(a generalization of the Kadyshevsky equation \cite{kadyshevsky})  where
\begin{eqnarray}
 {\cal M} &=& {\cal M}_{NN,NN} \oplus 
 \left(%
\begin{array}{cc}
{\cal M}_{\Lambda N,\Lambda N}  & {\cal M}_{\Lambda N,\Sigma N}  \\
{\cal M}_{\Sigma N,\Lambda N}   & {\cal M}_{\Sigma N,\Sigma N}  \end{array}%
\right) \oplus 
 \left(%
\begin{array}{cccc}
 {\cal M}_{\Lambda \Lambda,\Lambda \Lambda} & {\cal M}_{\Lambda \Lambda,\Xi N} & {\cal M}_{\Lambda \Lambda,\Sigma \Sigma} & {\cal M}_{\Lambda \Lambda,\Sigma \Lambda}\\
 {\cal M}_{\Xi N, \Lambda \Lambda} & {\cal M}_{\Xi N,\Xi N} & {\cal M}_{\Xi N,\Sigma\Sigma} & {\cal M}_{\Xi N,\Sigma\Lambda} \\
 {\cal M}_{\Sigma \Sigma,\Lambda \Lambda} & {\cal M}_{\Sigma \Sigma,\Xi N} & {\cal M}_{\Sigma \Sigma,\Sigma\Sigma} & {\cal M}_{\Sigma \Sigma,\Sigma \Lambda}\\
  {\cal M}_{\Sigma\Lambda,\Lambda\Lambda } & {\cal M}_{\Sigma\Lambda,\Xi N } & {\cal M}_{\Sigma\Lambda,\Sigma\Sigma } & {\cal M}_{\Sigma\Lambda,\Sigma\Lambda } \\
\end{array}%
\right) ,
\nonumber\\
V &=& V_{NN,NN} \oplus \left(%
\begin{array}{cc}
 V_{\Lambda N,\Lambda N}  & V_{\Lambda N,\Sigma N}   \\
 V_{\Sigma N,\Lambda N}   & V_{\Sigma N,\Sigma N} 
\end{array}%
\right)\oplus \left(%
\begin{array}{cccc}
V_{\Lambda \Lambda,\Lambda \Lambda} & V_{\Lambda \Lambda,\Xi N} & V_{\Lambda \Lambda,\Sigma \Sigma} & V_{\Lambda \Lambda,\Sigma \Lambda}\\
 V_{\Xi N, \Lambda \Lambda} & V_{\Xi N,\Xi N} & V_{\Xi N,\Sigma\Sigma} & V_{\Xi N,\Sigma\Lambda} \\
 V_{\Sigma \Sigma,\Lambda \Lambda} & V_{\Sigma \Sigma,\Xi N} & V_{\Sigma \Sigma,\Sigma\Sigma} & V_{\Sigma \Sigma,\Sigma \Lambda}\\
V_{\Sigma\Lambda,\Lambda\Lambda } & V_{\Sigma\Lambda,\Xi N } & V_{\Sigma\Lambda,\Sigma\Sigma } & V_{\Sigma\Lambda,\Sigma\Lambda } \\
\end{array}%
\right), \nonumber\\
G &=&   
G_{NN}  \oplus  G_{\Lambda N}   \oplus  G_{\Sigma N}  \oplus  G_{\Lambda \Lambda} \oplus  G_{\Xi N} \oplus  G_{\Sigma \Sigma}  \oplus  G_{\Sigma\Lambda} \,,
\label{defMatrixEq}
\end{eqnarray}
and the two-body Green functions read
\begin{equation}
G^{IJ}(E)= \frac{1}{\omega_I \omega_J}\, \frac{m_I m_J}{E - \omega_I-\omega_J+i \epsilon} \,,
\label{Gij}
\end{equation}
where $m_I$ and $\omega_I$ are  the mass and  energy of  the $I$-th baryon.

{ We calculate the BB scattering amplitude in
the center-of-mass system (CMS) and denote 
the three-momenta of the incoming and
outgoing baryons  by  $\vec p$ and  $\vec p\,'$, respectively.}
In the partial wave basis, Eq.~(\ref{Teq1}) leads to
the following coupled-channel equations with the partial wave projected potential
$V^{IJ,KL}_{l'l,s's,j}\left(p',p\right)$,
\begin{eqnarray}
T^{IJ,KL}_{l'l,s's,j}\left(E, {p'}
,p\right) &=& V^{IJ,KL}_{l'l,s's,j}\left(E,p',p\right) + 
\sum_{l'',s'',P,Q}
\int_0^\infty
\frac{d k\,k^2}{2 \pi^2} V^{IJ,PQ}_{l'l'',s's'' j}\left(E,p',k\right) G^{PQ}(E) \, T^{PQ,KL}_{l''l,s''s,j}
\left(E, k ,p\right),
\label{PWEHDR}
\end{eqnarray}
where $IJ, KL$ and $PQ$ denote initial, final and intermediate
particle channels, $l$, $l''$, $l'$ and $s$, $s''$, $s'$ correspond to their orbital
angular momentum and their spin, respectively, while $j$ refers to the total angular
momentum of BB states. 
A standard UV counting shows that in the limit of large
integration momenta, considered for the same potential, Eq.~(\ref{PWEHDR}) with the Green functions of Eq.~(\ref{Gij})
has a milder UV behaviour than the corresponding Lippmann-Schwinger equation.
Therefore, its solutions are expected to show less sensitivity to the variation of the
cutoff parameter.

\section{LO Baryon-baryon potential and renormalization} 
\label{Calculations}

\begin{figure}

\includegraphics[width=0.48\textwidth]{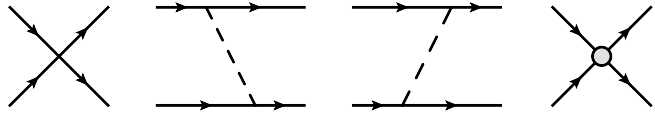}
\caption[]{\label{NNfig4:fig} Time-ordered diagrams contributing to
  the LO BB potential. 
The solid and dashed lines correspond to baryons and pseudoscalar mesons, respectively. 
 The contact interaction with filled blob represents the additional
 vertex due to the momentum-dependent pseudoscalar-meson--baryon
 interaction as specified in the diagrammatic rules (see the text).
  }
\end{figure}


We apply the standard Weinberg power counting to the BB
interaction potential, thus representing it as an 
expansion with a finite number of TOPT diagrams
at any given order. 
Diagrams contributing to the LO effective potential $ V_{\rm LO}$  for BB scattering are shown in
Fig.~\ref{NNfig4:fig}.  The potential   
consists of the short-range contact interaction part
$V_{0,C}^{IJ,KL}$, the long-range part generated by two
one-pseudoscalar-meson-exchange time-ordered 
diagrams and  the additional short-range contact term due to the
momentum-dependent interactions as explained in the diagrammatic rules.  
Adding together the  last three contributions (i.e. all but $V_{0,C}^{IJ,KL}$) and performing simplifications,  up to higher-order effects, we obtain the following expression
for the one-meson exchange contribution
\begin{eqnarray}
V_{0,M_P}^{IJ,KL} 
&=& -\frac{f_{IKP} f_{JLP} \,{\cal I}_{IJ,KL} }{2\, \omega(q,M_P) }\,
\left[ \frac{ 1}{\omega(q,M_P) +\omega(p_K,m_K) +\omega(p_J,m_J) -E-i\,\epsilon } \right. \nonumber\\
&&{}+ \left. \frac{1}{\omega(q,M_P) +\omega(p_L,m_L) +\omega(p_I,m_I) -E-i\,\epsilon } \right]  \frac{\left(m_I+m_K\right) \left(m_J+m_L\right)}{4
   \sqrt{m_I m_J m_K m_L}}
\nonumber\\
&&{}\times\frac{
   \left(\vec\sigma_1 \cdot\vec p_I \left(\omega
   \left(p_K,m_K\right)+m_K\right)- \vec\sigma_1 \cdot\vec p_K
   \left(\omega
   \left(p_I,m_I\right)+m_I\right)\right)
   }{
   \sqrt{\omega
   \left(p_I,m_I\right)+m_I}
   \sqrt{\omega \left(p_J,m_J\right)+m_J} \sqrt{\omega
   \left(p_K,m_K\right)+m_K} \sqrt{\omega \left(p_L,m_L\right)+m_L}} \nonumber\\
   &&{}\times  \left( \vec\sigma_2 \cdot\vec p_J  \left(\omega
   \left(p_L,m_L\right)+m_L\right)- \vec\sigma_2 \cdot\vec p_L
   \left(\omega \left(p_J,m_J\right)+m_J\right)\right)
 \,,
\label{OBEpot}
\end{eqnarray}
where $q=p_I-p_K =p_L-p_J$. 
  The isospin factors ${\cal I}_{IJ,KL} $ and the values of $f_{IKP}$ can be found in Refs.~\cite{Haidenbauer:2011ah,Haidenbauer:2007ra}.

The expressions for the LO contact interactions relevant for our  TOPT approach  ($V_{0,C}^{IJ,KL}$) are derived straightforwardly from the Lagrangian of Eq.(\ref{NNLagrdreg}) and 
can be found in Refs.~\cite{Ren:2016jna,Li:2016mln,Li:2018tbt}. 
It is important to emphasize that  both the 
one-meson exchange of Eq.(\ref{OBEpot}) and  the contact interaction potentials contain also higher-order contributions according to Weinberg's power counting. 
For example, for the contact interaction those emerge from the relativistic energy-dependent normalization factors of the nucleon spinors, as given by Eq.~(\ref{nuclspin}). 
 To single out LO contact interactions, we rewrite the relativistic expressions with baryon energies as 
\be \label{omegasplit}
\sqrt{ \omega
   \left(p,m\right)+m}=\sqrt{2 m}+\left[ \sqrt{ \omega
   \left(p,m\right)+m} -2 m\right]=\sqrt{2 m}+ {\cal O}(p^2/m), 
  \ee 
    and shift the term in the square brackets to the higher-order BB potential.  By doing so we obtain contact interactions which are identical to 
   those of the non-relativistic approach, see Refs.~\cite{Polinder:2006zh,Polinder:2007mp}.  We also perform such splitting for the terms in the numerator of  the expression in Eq.~(\ref{OBEpot}), that is 
   we replace  the particle energies  by their masses up to  higher-order effects.   However, in full analogy to the treatment of the Green function (see Eq.~(\ref{Gij})), 
   we keep the full energy expressions in the denominator in Eq.~(\ref{OBEpot}), 
   because performing their non-relativistic expansion is not commutative with loop integration \cite{Epelbaum:2012ua}. 
  While after renormalization  relativistic effects as such are not expected to play a significant role at low energies, keeping the full unexpanded expressions in the denominators 
  allows one to carry out  renormalization of the LO BB amplitude by obtaining cutoff  independent results -- see further discussion below.   
 Thus, for the  one-meson-exchange contribution to the LO potential we finally obtain
 \begin{eqnarray}\label{OBEpotLO}
V_{{\rm LO},M_P}^{IJ,KL} &=& -\frac{f_{IKP} f_{JLP} \,{\cal I}_{IJ,KL} }{2\, \omega(q,M_P) }\,
\left[ \frac{ 1}{\omega(q,M_P) +\omega(p_K,m_K) +\omega(p_J,m_J) -E-i\,\epsilon } \right. \nonumber\\
&&{}+ \left. \frac{1}{\omega(q,M_P) +\omega(p_L,m_L) +\omega(p_I,m_I) -E-i\,\epsilon } \right]  \frac{\left(m_I+m_K\right) \left(m_J+m_L\right)}{
   \sqrt{m_I m_J m_K m_L}}
\\
&&{}\times\frac{
   \left(m_K \vec\sigma_1 \cdot\vec p_I - m_I \vec\sigma_1 \cdot\vec p_K
  \right){}  \left( m_L \vec\sigma_2 \cdot\vec p_J  - m_J\vec\sigma_2 \cdot\vec p_L \right)
   }{
   \sqrt{\omega
   \left(p_I,m_I\right)+m_I}\,
   \sqrt{\omega \left(p_J,m_J\right)+m_J}\, \sqrt{\omega
   \left(p_K,m_K\right)+m_K}\, \sqrt{\omega \left(p_L,m_L\right)+m_L}} . \nonumber
\end{eqnarray}
The difference between $V_{0,M_P}^{IJ,KL}$ and $V_{{\rm LO},M_P}^{IJ,KL}$ (cf. Eqs.~(\ref{OBEpot}) and (\ref{OBEpotLO})) is
included in the higher-order corrections.   The resulting LO long-range potential has a milder ultraviolet
behaviour than its analog obtained by using the LO approximation for Dirac spinors as done in Ref.~\cite{Epelbaum:2012ua}. 
A distinctive feature of the potential of Eq.~(\ref{OBEpotLO}) as compared to its  non-relativistic analog
is that its iterations within  the integral equations (\ref{PWEHDR}) lead to ultraviolet finite diagrams. 
To demonstrate this feature, consider e.g.~the one-loop integral
\begin{equation}
 \sum_{Q,R}
\int
\frac{d^3\vec k}{(2 \pi)^3} X_{VGV} \equiv \sum_{Q,R}
\int \frac{d^3\vec k}{(2 \pi)^3} V^{IJ,QR}_{{\rm LO},M_P}\,  G^{QR}(E)\,  V^{QR,KL}_{{\rm LO},M_P}\,,
\label{1loop}
\end{equation}
where $\vec p_Q=\vec p_I - \vec k $ and $\vec p_R=\vec p_J + \vec k $.
For $k \equiv |\vec k| \to\infty$, we obtain for the integrand
\begin{eqnarray}
X_{VGV} &=& \sum_{Q,R} f_{IQP} f_{JRP} \,{\cal I}_{IJ,QR} f_{QKP} f_{RLP} \,{\cal I}_{QR,KL}  \,\frac{\left(m_I+m_Q\right) \left(m_J+m_R\right) \left(m_Q+m_K\right) \left(m_R+m_L\right)}{4\, m_Q m_R
   \sqrt{m_I m_J m_K m_L}}\nonumber\\
&\times &
\frac{
   \left( m_Q \, \vec\sigma_1 \cdot\vec p_I  - m_I\, \vec\sigma_1 \cdot (\vec p_I-\vec k) \right) 
   \left(  m_K \,  \vec\sigma_1 \cdot (\vec p_I-\vec k ) - m_Q\, \vec\sigma_1 \cdot \vec p_K)
   \right) 
   }{
  k^3 \sqrt{\omega   \left(p_I,m_I\right)+m_I} \,
   \sqrt{\omega \left(p_J,m_J\right)+m_J} 
   }\,
   \frac{(-m_Q m_R)}{2 k^3} \nonumber\\
   &\times&  \frac{
    \left( m_R\, \vec\sigma_2 \cdot\vec p_J  - m_J\, \vec\sigma_2 \cdot(\vec p_J+\vec k)
   \right) \left( m_L\, \vec\sigma_2 \cdot (\vec p_J+\vec k ) - m_R\, \vec\sigma_2 \cdot\vec p_L
    \right)
   }{
k^3   \sqrt{\omega
   \left(p_K,m_K\right)+m_K}\, \sqrt{\omega \left(p_L,m_L\right)+m_L}} \,,
\label{1loopUV}
\end{eqnarray}
where two factors of $1/k^3$ stem from the denominators of the one-meson exchange potentials of Eq.~\eqref{OBEpotLO} and an additional factor of
$1/k^3$ represents  the UV behavior of the Green function of  Eq.\eqref{Gij}. 
The integrand $X_{VGV}$ behaves as $\sim 1/k^5$ and thus leads to an UV convergent one-loop integral. 
Analogously, it can be easily shown that all iterations of the one
pseudoscalar meson-exchange potential lead to  UV finite diagrams.
Then,   the full LO potential, which also contains the contact interactions, is
perturbatively renormalizable since all divergences appearing from
its iterations can be absorbed in the coupling constant of the contact interaction.  As a consequence,  the ultraviolet cutoff 
can be safely removed (set to infinity) at LO which allows one to avoid  
finite-cutoff artefacts inherent to   the conventional non-relativistic framework.
Also, in this approach, one does not face  a well-known issue of the integral equation having non-unique solution for singular attractive potentials 
(this is e.g. the case for the sufficiently strong $1/r^2$ attractive potential substituted in the Lippmann-Schwinger equation)  -- see, e.g., 
 Ref.~\cite{Epelbaum:2012ua} for more details. 
For our LO potential the 
 integral equations (\ref{PWEHDR}) for the BB scattering amplitudes have unique solutions for all partial waves.  


\medskip

Although the LO potential can be properly renormalized as discussed above,  in certain channels of BB scattering one may still run into the situation that
corrections beyond LO are large enough to require their  nonperturbative treatment. Below, we address this issue in detail on
the example of NN scattering.

The LO  NN potential consists of two momentum-independent
contact interactions contributing to  S-waves and the one-pion
exchange potential (OPEP) corresponding to Eq.~(\ref{OBEpotLO}), which has the
form 
\begin{equation}
-\frac{g_A^2}{4\,F_0^2}\,\frac{\
\vec \tau_1\cdot\vec\tau_2}{\omega(p-p\,',M_\pi)} \frac{4 m_N^2}{(m_N+\omega(p,m_N) )(m_N+ \omega(p',m_N) )} 
\frac{\left[
\vec\sigma_1\cdot \left(\vec p-\vec p\,'\right)\right]\left[
\vec\sigma_2\cdot \left(\vec p-\vec p\,'\right)\right]}{ \omega(p-p\,',M_\pi) +\omega(p,m_N) + \omega(p',m_N)  
-E-i\,\epsilon },
\label{opepotreduced}
\end{equation}
where $E=2 \sqrt{m_N^2+q_{\rm on}^2}$ with $q_{\rm on}$ being the absolute value of
the three-momentum of  
nucleons in the
center-of-mass frame. The LO potential, substituted in the
equations of (\ref{PWEHDR}),  
is perturbatively renormalizable and the integral equations have
unique solutions. 
As a general trend,  the calculations   done at LO
 provide a reasonable description of the empirical phase
shifts as shown in Fig.~\ref{oneloop:fig}.
\begin{figure}
\includegraphics[width=\textwidth]{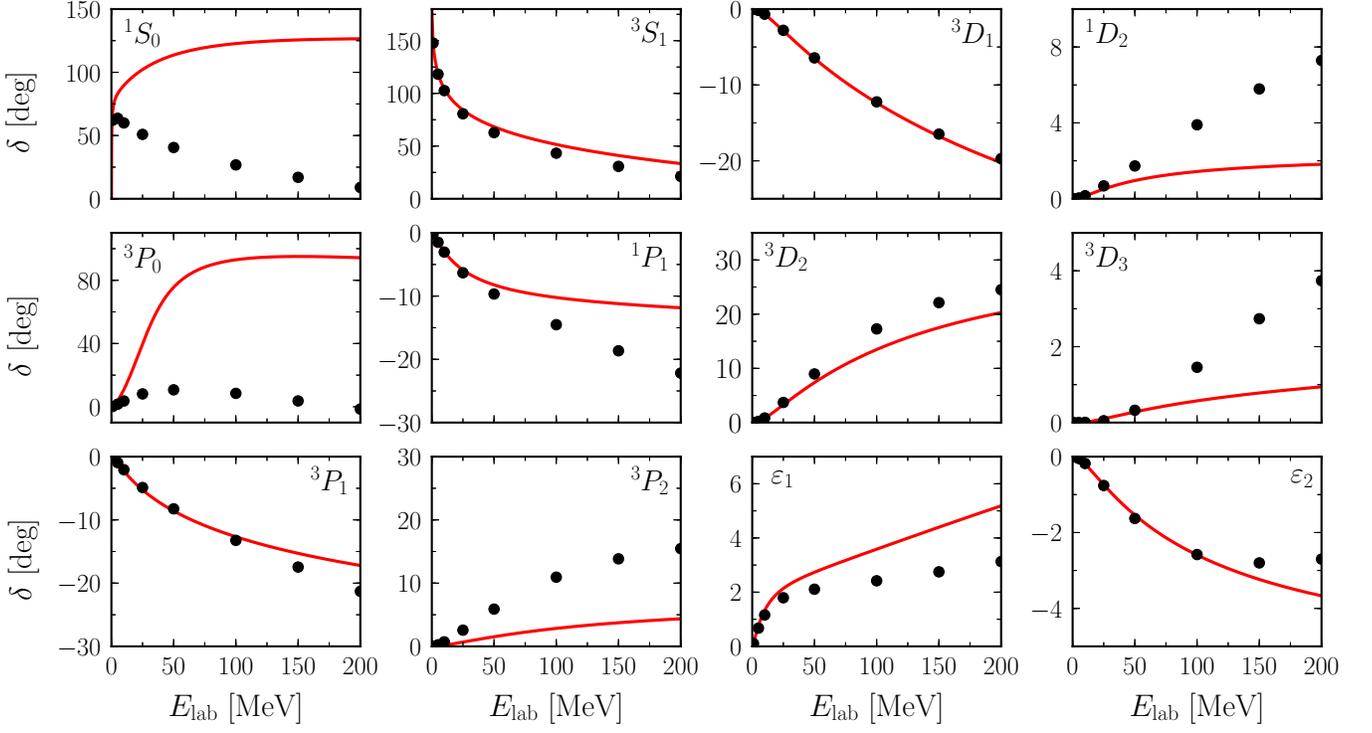}
\caption[]{\label{oneloop:fig} Nucleon-nucleon phase shifts and mixing
  angles at LO. Lines correspond to our results while 
  solid dots refer to the Nijmegen partial wave analysis \cite{Stoks:1993tb}.}
\end{figure}
However, one recognizes
large deviations from the Nijmegen partial wave analysis (PWA) for the
cases of the $^1$S$_0$ and $^3$P$_0$ partial waves. In the  $^1$S$_0$
channel, the observed discrepancy can be traced back to
the large (as compared to the inverse range of the OPEP) experimental value of the
effective range. In the $^3$P$_0$ channel, the OPEP is known to become
non-perturbative at rather low momenta \cite{Birse:2005um}. 
 Given that in this channel  the
loop integrals in our Lorentz-invariant formulation are
effectively cut off at momenta of the order of the nucleon mass and there is no contact interaction at LO, the
observed discrepancy does not come as a surprise. 
For both the  $^1$S$_0$ and $^3$P$_0$ partial waves, the large differences between
the LO results and the empirical phase shifts suggest that at least
a part of the subleading corrections must be treated
non-perturbatively. While the non-perturbative inclusion of
pion-exchange potentials beyond LO would generally destroy the explicit
renormalizability feature of our approach and thus prevent one from
eliminating the cutoff, it is still possible to
treat the sub-leading contact interactions non-perturbatively within a
cutoff-independent approach in the way consistent with the principles
of EFT.  For the $^1$S$_0$ channel, it was already 
demonstrated in Ref.~\cite{Epelbaum:2015sha} that  the proper inclusion of the NLO contact interactions in the considered non-perturbative
approach results in the significant improvement for the phase shift, see also Ref.~\cite{Baru:2015ira} for related discussion in the context of low-energy theorems. The use of the modified OPE potential from Eq.~(\ref{opepotreduced}) in the $^1S_0$ case is expected to produce results similar to those of Ref.~\cite{Epelbaum:2015sha}.

In the following, we consider in detail 
the   $^3$P$_0$ partial wave. 
 To improve the description of the $^3$P$_0$ phase shift, we  
add the lowest-order contact interaction term to the potential which is
treated non-perturbatively thus obtaining 
\bea
\label{LO}
V_{\rm LO}^{^3P_0}\left(p\,',p \, \right)  =  C\,p\,' p
+  V_\pi
\equiv V_C + V_\pi ,
\eea
where $V_\pi $ stands for the OPEP of Eq.~(\ref{opepotreduced}) projected onto the $^3P_0$ partial wave.
For the above potential it is possible to write the solution to
the integral equation in such a form (analogously to Ref.~\cite{Kaplan:1996xu}),
which allows one to carry out the subtractive renormalization.
{ For this purpose, we write the integral equations \eqref{PWEHDR} symbolically as 
\begin{equation}
T=V+V\,G\,T,
\label{eqsim}
\end{equation}
and, analogously to Ref.~\cite{Epelbaum:2015sha}, present their solution, 
for a separable contact interaction potential
\begin{equation}
V_C(p',p)= \bar\xi(p')\, {\cal C}\, \xi(p),
\label{nuCfact}
\end{equation}
as 
\begin{equation}
T=T_\pi+(1+T_\pi\,G)\,\bar\xi\,{\cal X}\,\xi (1+G\,T_\pi),
\label{taup}
\end{equation}
where 
\begin{equation}
{\cal X}= \left[{\cal C}^{-1}-\xi\,G\,\bar\xi -  \xi\,G\,T_\pi G\,\bar\xi \,\right]^{-1},
\label{chisol}
\end{equation}
and the amplitude $T_\pi$ satisfies the equation
\begin{equation}
T_\pi=V_\pi+V_\pi\,G\,T_\pi\,.
\label{OEQ1}
\end{equation}
}
In a close analogy to Ref.~\cite{Epelbaum:2015sha}, we apply the
subtractive (BPHZ-type) renormalization, i.e. we subtract {\it all} divergences in loop diagrams 
 and  replace the coupling constants  by their renormalized, finite values (see, e.g.,  Ref.~\cite{Collins:1984xc} for further details of BPHZ renormalization).  
Subtractive renormalization of the
considered problem corresponds to the inclusion of contributions of an infinite number of counter terms generated by bare parameters of the  effective Lagrangian \cite{Epelbaum:2018zli}.

We need to apply subtractive renormalization to the expression of Eq.~(\ref{taup}), where  for $V_C$ from Eq.~\eqref{LO} 
we have
\begin{equation}
{\cal C} = 
 C, \ \ \
\bar\xi(p\,')  =    p\,' ,\ \ \
\xi(p) = p\,.
\label{cxi}
\end{equation}
Furthermore,  by analysing the asymptotic behavior of  the OPEP of Eq.\eqref{opepotreduced}  in  the $^3$P$_0$ channel,  
\be \label{OPEas}
V_\pi(p',p)\big|_{\scriptscriptstyle p\to\infty,\,  p'<\infty}
  \sim \frac{1}{ p^2}\,, \quad 
  V_\pi(p',p)\big|_{\scriptscriptstyle p<\infty,\, p'\to\infty }
  \sim \frac{1}{ p'^2}\,, \quad  
  V_\pi(p',p)\big|_{\scriptscriptstyle p\to\infty,\,p'\to\infty}\sim \frac{1}{p\, p'}\,,
\ee
one is led to conclude that the amplitude $T_\pi$ is finite, and so are  $\bar\Xi(p')=(1+T_\pi\,G)\,\bar\xi$  and $\Xi(p)=\xi (1+G\,T_\pi)$. 
All (multi-loop)
sub-diagrams contained in $\xi\,G\,T_\pi G\,\bar\xi$ are also finite,
so that this quantity contains only the overall logarithmic divergence
which stems from the regime
$p\to\infty,\,p'\to\infty$
in Eq. \eqref{OPEas}.  
On the other hand, the term  $\xi\,G\,\bar\xi $ is quadratically divergent and therefore 
requires additional BPHZ subtractions.  Indeed,  in the limit when the cutoff 
$\Lambda$  is much larger than the nucleon mass 
$\xi\,G\,\bar\xi $  can be written  as
\be \label{loopdiv}
\xi\,G\,\bar\xi =  -\frac{m_N^2}{8\pi^2}\biggl( \Lambda^2 + E\,  \Lambda + (3m_N^2 - E^2/2) \log(\Lambda/m_N) + f(E)\biggr) , 
\ee
where $f(E)$ is a finite function.  Applying  the subtractive renormalization, the final   
 renormalized expression reads
\begin{equation}
T(p\,',p)=T_\pi(p\,',p)+\frac{ \bar \Xi(p\,')\,\Xi(p)}{\frac{1}{C_R} - (\xi\,G\,\bar\xi)^R - (\xi\,G\,T_\pi G\,\bar\xi-\alpha)}\,,
\label{tRen}
\end{equation}
where the 
$\alpha$ subtracts the overall divergence of $\xi\,G\,T_\pi G\,\bar\xi$. The subtracted expression of $(\xi\,G\,\bar\xi)^R$ is given in Ref.~\cite{Epelbaum:2015sha}
\begin{eqnarray}
(\xi\,G\,\bar\xi)^R &=& q_{\rm on}^2 I_0^R(\mu,q_{\rm on})=\frac{m_N^2 q_{\rm on}^2}{4\pi^2 E}\left(2q_{\rm on}\left( \,\sinh^{-1}\frac{q_{\rm on}}{m_N}-i\pi\right)-\pi m_N\right)\nonumber\\
&+&\frac{m_N^2q_{\rm on}^2}{8\pi^2\sqrt{m_N^2-\mu^2}}\left(2\mu\left( \,\sin^{-1}\frac{\mu}{m_N}-\pi\right)+\pi m_N\right)\,,
\label{GchiiR}
\end{eqnarray}
where $\mu$ is the renormalization scale.
In practice, we fix the bare constant $1/C=1/C_R+\alpha$ as a function of  the
cutoff numerically in such a way that it cancels the divergent part of
$\xi\,G\,T_\pi G\,\bar\xi$ and the resulting cutoff-independent
scattering amplitude describes the phase shift for a fixed value of
the energy chosen to be 20 MeV.  
The resulting NN phase shift in the $^3P_0$ channel  is plotted in Fig.~\ref{3P0Per:fig} (right panel) and shows a very good description of the data.  The renormalized amplitude shows weak dependence on the
renormalization { scale $\mu$ - the variation } of $\mu$ from the
pion mass to the nucleon mass results in an effect of
$\sim 0.6$ degrees
in the phase shift for $E_{\rm lab}\sim$ 100 MeV.

\begin{figure}
\includegraphics[width=0.48\textwidth]{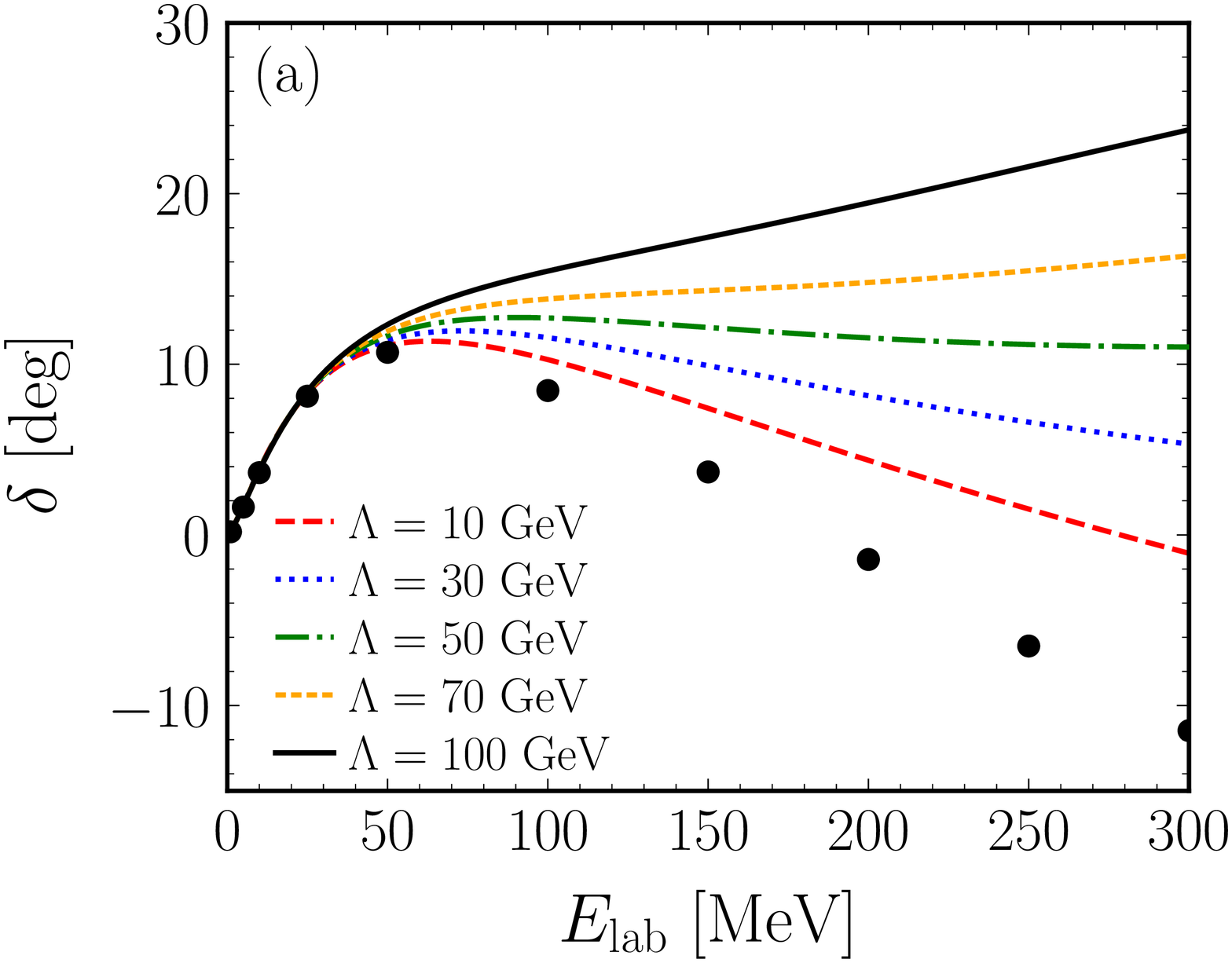}
\hfill
\includegraphics[width=0.48\textwidth]{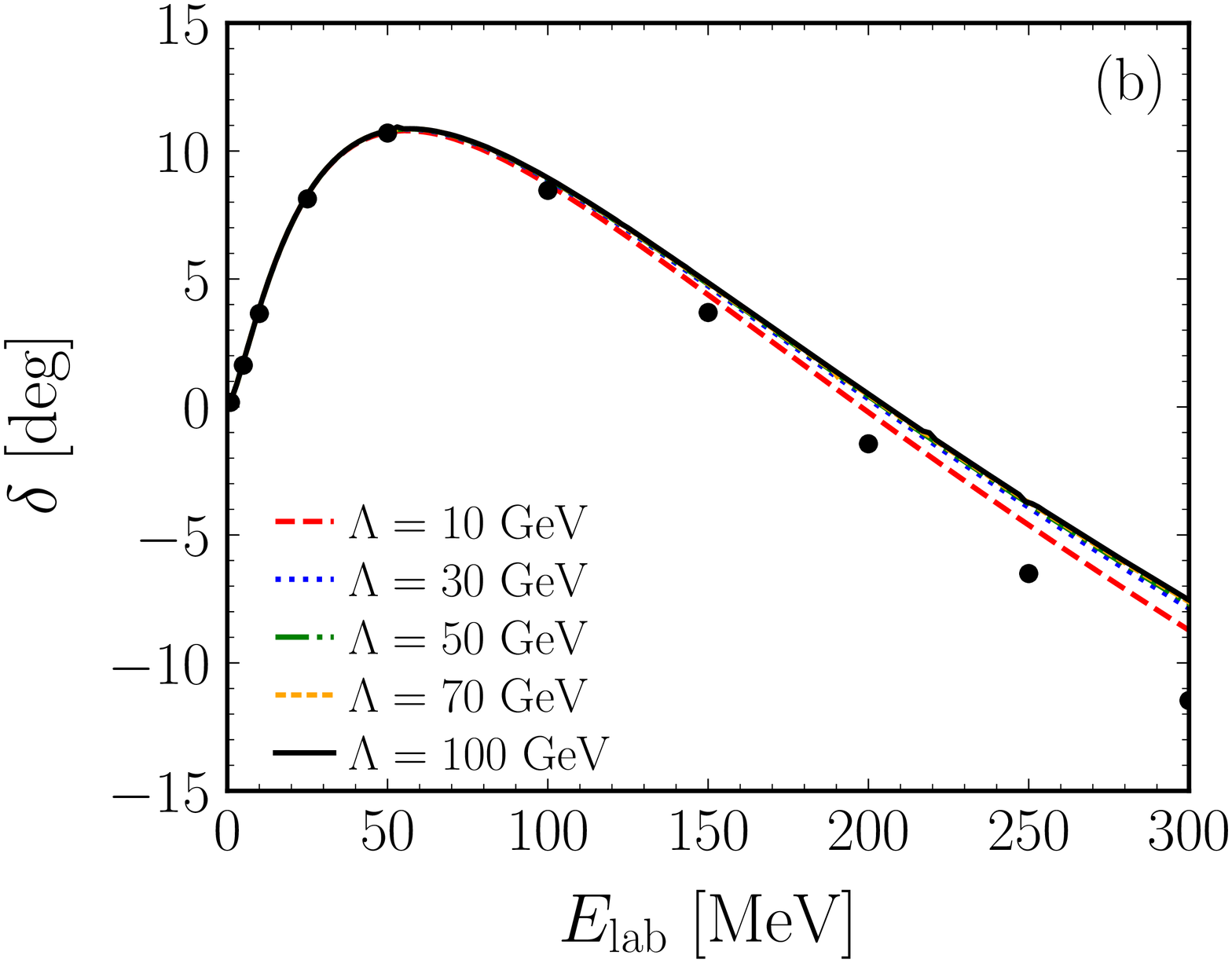}
\caption[]{\label{3P0Per:fig} Cutoff-dependence of the calculated
  $^3P_0$ phase shifts.  Left and right panels correspond to
  the ``non-perturbatively renormalized" and subtractively renormalized
  amplitudes, respectively.  
Solid dots are the results of the Nijmegen PWA \cite{Stoks:1993tb}.}
\end{figure}


  It is instructive to compare the above subtractive
  renormalization with the so-called ``non-perturbative
  renormalization", where the contact interaction is tuned to
  reproduce the empirical value of the scattering amplitude at a given
  energy {\it without subtracting all ultraviolet divergences} as
  advocated e.g.~in
  Refs.~\cite{Nogga:2005hy,PavonValderrama:2005uj,Long:2011qx}
  within the nonrelativistic framework.
Such  ``non-perturbative renormalization" does remove the quadratic divergence
  in $\xi\,G\,\bar\xi $ and the logarithmic divergence in
  $\xi\,G\,T_\pi G\,\bar\xi$  with the energy-independent prefactors, while the linear and logarithmic divergences 
  with energy-dependent coefficients still survive in $\xi\,G\,\bar\xi$ (see Eq. \eqref{loopdiv}).
In Fig.~\ref{3P0Per:fig} we confront the cutoff
dependence of the $^3$P$_0$ phase shift for such
a ``non-perturbatively renormalized" scattering amplitude (left panel) with the
residual cutoff dependence in the properly renormalized approach as
discussed above (right panel). 
In line with the reasoning discussed above, one may  conclude that 
the 
approach without explicit subtractions of divergences does
not lead to a properly renormalized result for the scattering amplitude. 
{ This conclusion is in full agreement with Refs.~\cite{Epelbaum:2009sd,Epelbaum:2018zli}.  }

\section{Summary}
\label{conclusions}

In this paper we considered the baryon-baryon (BB) scattering problem in the framework
of manifestly Lorentz-invariant formulation of SU(3) BChPT  by 
applying time-ordered perturbation theory. 
By integrating over zeroth components of loop momenta in Feynman
diagrams we formulated the diagrammatic rules of time-ordered
perturbation theory, which can be applied to momentum-dependent
interactions and particles with non-zero spin. For the case of
BB scattering, the importance of time-ordered diagrams can
be determined using the
Weinberg's power counting rules \cite{Weinberg:rz,Weinberg:um}.
An infinite number of diagrams contributes to the BB
scattering amplitude at any finite order. To sum up the relevant
contributions it is convenient to define the effective potential
as a sum of all two-baryon irreducible contributions to the scattering
amplitude within TOPT.
In a full analogy with the conventional nonrelativistic framework,
the scattering amplitudes are obtained as solutions to a system of coupled-channel
integral equations with the potentials at the corresponding order. 
These equations represent a coupled-channel generalization of the
Kadyshevsky equation 
\cite{kadyshevsky} and feature a milder ultraviolet behaviour as compared to
their non-relativistic analogs.  We obtained new perturbatively
  renormalizable LO BB potential  which leads to unique
  solutions of the integral equations for scattering amplitudes in all
  partial waves. 
On the example of NN scattering we addressed the
issue of the non-perturbative inclusion of the leading short-range
interaction in the  $^3P_0$ partial wave.
For this purpose,  we  carried out subtractive  renormalization
in a way consistent with EFT.  We also considered the
``non-perturbative renormalization" approach for the problem at hand as
advocated, e.g.,~in Refs.~\cite{Nogga:2005hy,PavonValderrama:2005uj,Long:2011qx} 
to determine the value of the contact interaction. 
The resulting cutoff dependence of the amplitude supports
the conclusions of Refs.~\cite{Epelbaum:2009sd,Epelbaum:2018zli}
about the incompatibility of  
such ``non-perturbative renormalization" with the principles of EFT.

The established formalism can be used to study BB
scattering in the SU(3) sector based on a renormalizable formulation
with the corrections beyond LO treated perturbatively. Such a
framework permits a complete removal of the ultraviolet cutoff
$\Lambda$ by taking the limit $\Lambda \to \infty$, see Refs.~\cite{Epelbaum:2012ua,Epelbaum:2013ij,Epelbaum:2013naa}
for applications in the non-strange sector. Alternatively, one may
follow a more traditional approach by solving the integral equations
for a truncated potential without relying on a perturbative
treatment of higher-order contributions as it
is usually done for NN scattering, see
e.g.~\cite{Epelbaum:2014efa,Epelbaum:2014sza,Reinert:2017usi,Entem:2017gor}.
In that case
the  $\Lambda \to \infty$ limit is not legitimate anymore, but one 
may still expect to benefit from the milder ultraviolet behavior of
the integral equations in the Lorentz-invariant formulation, which 
should provide more flexibility in the choice of the ultraviolet
cutoff. Work along these lines is in progress. 

\section*{Acknowledgments}
We are grateful to Ulf-G.~Mei{\ss}ner for useful comments on the
manuscript. 
This work was supported in part by the Georgian Shota Rustaveli National
Science Foundation (Grant No. FR17-354),  by DFG and NSFC
 through funds provided to the
Sino-German CRC 110 ``Symmetries and the Emergence of Structure in QCD" (NSFC
Grant No.~11621131001, DFG Grant No.~TRR110), the BMBF  (Grant
No.~05P18PCFP1) and the Russian Science Foundation (Grant
No.~18-12-00226).

\end{document}